\documentclass[trackchanges]{aastex701}

\newcommand{\Dec}{$-2.575^{+0.002}_{-0.002}$}
\newcommand{\Distance}{$109^{+6}_{-8}\ \mathrm{kpc}$}
\newcommand{\DistanceModulus}{$20.16^{+0.03}_{-0.03}$}
\newcommand{\Ellipticity}{$< 0.61$}
\newcommand{\ExtensionRadialArcmin}{$r_h = 0.60^{+0.14}_{-0.17}$}

\newcommand{\Iau}{J2201$-$0234}
\newcommand{\MvMartin}{$M_V=-1.9^{+0.6}_{-1.0}$}

\newcommand{\PhysicalSizeRadial}{$r_{1/2} = 19^{+4}_{-6}$}

\newcommand{\Ra}{$330.474^{+0.002}_{-0.002}$}

 \usepackage{enumitem}
 \usepackage{float}

\usepackage{amsmath}
\usepackage{amssymb}
\usepackage{xspace}
\usepackage{xifthen}
\usepackage{eso-pic}

\definecolor{forestgreen}{HTML}{228B22}
\definecolor{urlblue}{HTML}{000000}

\mathchardef\mhyphen="2D

\newlength{\dhatheight}

\newcommand{\unit}[1]{\ensuremath{\mathrm{\,#1}}\xspace}

\newcommand{\pc}{\unit{pc}}

\newcommand{\e}{\unit{e^{-}}}

\newcommand{\figref}[1]{Figure~\ref{fig:#1}}

\newcommand{\bandvar}[2][]{%
  \ifthenelse{\isempty{#1}}{\var{#2}}{\var{#2\_#1}}%
}

\newcommand{\ra}{{\ensuremath{\alpha_{2000}}}\xspace}
\newcommand{\dec}{{\ensuremath{\delta_{2000}}}\xspace}

\newcommand{\var}[1]{\ensuremath{\texttt{\MakeUppercase{#1}}}\xspace}

\providecommand\physrep{\ref@jnl{Phys.~Rep.}}%
\providecommand\apjs{\ref@jnl{ApJS}}%
\providecommand{\jcap}{\ref@jnl{JCAP}}%
\shorttitle{Aquarius~IV: Rubin's First Milky Way Satellite Discovery}

\shortauthors{Cerny, Pai, Drlica-Wagner et al.}

\begin{document}

\title{Discovery of the Distant, Ultra-Faint Milky Way Satellite Aquarius~IV \\\ with the Vera C. Rubin Observatory Early Data Preview 2}

\author[0000-0003-1697-7062]{William Cerny}
\affiliation{Department of Astronomy, Yale University, New Haven, CT 06520, USA}
\email{william.cerny@yale.edu}

\author[0009-0008-9641-6065]{Aashay Pai}
\email{pai@uchicago.edu}
\affiliation{Department of Physics, University of Chicago, Chicago, IL 60637, USA}
\affiliation{Kavli Institute for Cosmological Physics, University of Chicago, Chicago, IL 60637, USA}
\affiliation{NSF-Simons AI Institute for the Sky (SkAI), 172 E. Chestnut St., Chicago, IL 60611, USA}

\author[0000-0001-8251-933X]{Alex Drlica-Wagner}
\email{kadrlica@uchicago.edu}
\affiliation{Fermi National Accelerator Laboratory, P. O. Box 500, Batavia, IL 60510, USA}
\affiliation{Kavli Institute for Cosmological Physics, University of Chicago, Chicago, IL 60637, USA}
\affiliation{Department of Astronomy and Astrophysics, University of Chicago, Chicago, IL 60637, USA}
\affiliation{NSF-Simons AI Institute for the Sky (SkAI), 172 E. Chestnut St., Chicago, IL 60611, USA}

\author[0000-0002-6021-8760]{Andrew~B.~Pace}
\email{pvpace1@gmail.com}
\affiliation{Department of Astronomy, University of Virginia, 530 McCormick Road, Charlottesville, VA 22904, USA}

\author[0000-0001-6957-1627]{Peter S. Ferguson}
\affiliation{Department of Astronomy and DiRAC Institute, University of Washington, 3910 15th Ave NE, Seattle, WA, 98195, USA}
\email{pferguso@uw.edu}

\author[0000-0002-7007-9725]{Marla~Geha}
\email{marla.geha@yale.edu}
\affiliation{Department of Astronomy, Yale University, New Haven, CT 06520, USA}

\author[0000-0003-0478-0473]{Chin~Yi~Tan}
\email{chinyi@uchicago.edu}
\affiliation{Kavli Institute for Cosmological Physics, University of Chicago, Chicago, IL 60637, USA}
\affiliation{Department of Physics, University of Chicago, Chicago, IL 60637, USA}
\affiliation{NSF-Simons AI Institute for the Sky (SkAI), 172 E. Chestnut St., Chicago, IL 60611, USA}

%%%%%%%%%%%%%%%%%%%%%%%%%%%%%%%%%%%%%%%%%%%%%%%%%%%
 %ALPHABETICAL HEREAFTER
\author[0009-0007-9488-7050]{Sasha Campana}
\affiliation{Department of Physics and Astronomy, Dartmouth College, Hanover, NH 03755, USA}
\email{sasha.n.campana.gr@dartmouth.edu}

\author[0000-0002-3936-9628]{Jeffrey L. Carlin}
\affiliation{NSF–DOE Vera C. Rubin Observatory/NSF NOIRLab, 950 N. Cherry Ave., Tucson, AZ 85719, USA}
\email{jeff.carlin@noirlab.edu}

\author[0000-0002-1763-4128]{Denija Crnojevi\'{c}}
\affiliation{Department of Physics \& Astronomy, University of Tampa, 401 West Kennedy Boulevard, Tampa, FL 33606, USA}
\email{dcrnojevic@ut.edu}

\author[0000-0002-4863-8842]{Alexander~P.~Ji}
\affiliation{Department of Astronomy and Astrophysics, University of Chicago, Chicago, IL 60637, USA}
\affiliation{Kavli Institute for Cosmological Physics, University of Chicago, Chicago, IL 60637, USA}
\affiliation{NSF-Simons AI Institute for the Sky (SkAI), 172 E. Chestnut St., Chicago, IL 60611, USA}
\email{alexji@uchicago.edu}

\author[0000-0002-9269-8287]{Guilherme~Limberg}
\affiliation{Kavli Institute for Cosmological Physics, University of Chicago, Chicago, IL 60637, USA}
\affiliation{Department of Astronomy and Astrophysics, University of Chicago, Chicago, IL 60637, USA}
\email{limberg@uchicago.edu}

\author[0000-0002-8093-7471]{Pol Massana}
\affiliation{NSF NOIRLab, Casilla 603, La Serena, Chile}
\email{pol.massana@noirlab.edu}

\author[0000-0003-3519-4004]{Sidney Mau}
\affiliation{Department of Physics, Duke University, Durham, NC 27708, USA}
\email{sidney.mau@duke.edu}

\author[0000-0003-0105-9576]{Gustavo E. Medina}
\affiliation{David A. Dunlap Department of Astronomy \& Astrophysics, University of Toronto, 50 St George Street, Toronto ON M5S 3H4, Canada}
\affiliation{Dunlap Institute for Astronomy \& Astrophysics, University of Toronto, 50 St George Street, Toronto, ON M5S 3H4, Canada}
\email{gustavo.medina@astro.utoronto.ca}

\author[0000-0001-9649-4815]{Bur{\c{c}}in Mutlu-Pakdil}
\affiliation{Department of Physics and Astronomy, Dartmouth College, Hanover, NH 03755, USA}
\email{Burcin.Mutlu-Pakdil@dartmouth.edu}

\author[0000-0002-1594-1466]{Joanna D. Sakowska}
\affiliation{Instituto de Astrofísica de Andalucía (IAA-CSIC), Glorieta de la Astronom\'ia s/n, E-18008 Granada, Spain}
\email{jsakowska@iaa.csic.es}

\author[0000-0003-2497-091X]{Nora Shipp}
\affiliation{Department of Astronomy and DiRAC Institute, University of Washington, 3910 15th Ave NE, Seattle, WA, 98195, USA}
\email{nshipp@uw.edu}

\author[0000-0003-1479-3059]{Guy~S.~Stringfellow}
\affiliation{University of Colorado Boulder, Boulder, CO 80309, USA}
\email{Guy.Stringfellow@colorado.edu}

\correspondingauthor{ \\ William Cerny (\url{william.cerny@yale.edu})}

\begin{abstract}
We present the discovery of Aquarius~IV (Rubin\,J2201$-$0234) -- the first ultra-faint Milky Way satellite to be identified using data from the Vera C.\ Rubin Observatory.  This system was detected at $\sim$8$\sigma$ significance using Rubin Early Data Preview 2 (EDP2) photometry and independently confirmed at $\sim$6$\sigma$ significance in archival Dark Energy Camera imaging. Jointly fitting its morphology and distance,  we find that Aquarius~IV is a low-luminosity (\MvMartin), compact (\PhysicalSizeRadial~pc; \ExtensionRadialArcmin~arcmin) stellar system in the outer Galactic halo ($D_{\odot} = $~\Distance). Its stellar population is consistent with an ancient, metal-poor stellar isochrone ($\tau = 13$~Gyr, $Z=0.0001$). These properties closely resemble those of the smallest and faintest confirmed ultra-faint dwarf galaxies, though a globular cluster classification is not ruled out. Given the small number of detected member stars in Rubin EDP2, deeper imaging and spectroscopy will be critical for determining the properties and classification of Aquarius~IV at higher confidence. 
\end{abstract}

\section{Introduction}
Ultra-faint dwarf galaxies (UFDs) are the least luminous, least chemically enriched, and most dark-matter-dominated galaxies known, making them powerful probes of galaxy formation and dark matter physics \citep{2019ARA&A..57..375S}. Their extremely low luminosities ($L_\star \lesssim 10^5\,L_\odot$) and diffuse stellar distributions make them challenging to detect \citep{2010AdAst2010E..21W}, even around the Milky Way (MW). In the two decades since the first MW UFD was discovered, the known population has increased to $>$50. However, the census of UFDs around the MW remains far from complete, with potentially hundreds left to discover \citep[][]{Nadler:2020}.

\par The NSF-DOE Vera C.\ Rubin Observatory (hereafter, Rubin) began the ten-year Legacy Survey of Space and Time (LSST) in June 2026. A key objective of Rubin LSST is to map the MW halo, revealing previously-hidden satellites, stellar streams, and other resolved stellar substructures \citep{2019ApJ...873..111I}. 
To enable science from pre-survey observations, $\sim$3000 deg$^2$ of data taken between April 2025 and January 2026 were released on 27 July 2026 as part of Early Data Preview 2 \citep[EDP2;][]{RTN-115,DP2reference}. Here, we present the first discovery of a UFD candidate in Rubin data from an initial search of EDP2.

\section{Search and Discovery}
We searched for UFDs in EDP2 with a Rubin-compatible implementation of the \texttt{simple} algorithm \citep{Bechtol15, 2025OJAp....8E..89T}. We constructed a stellar sample by querying the EDP2 \texttt{Object} table with LSDB \citep{2025arXiv250102103C} and selecting stars (\texttt{refExtendedness} = 0) with blue colors ($g-r < 1$), low extinction ($E(B-V) < 1$), and PSF magnitudes brighter than $g = 25$, $r=24.5$, $i=24.5$. The EDP2 magnitudes were corrected for interstellar extinction following \citet{2026arXiv260725044P}; hereafter, extinction-corrected magnitudes are denoted with a ``0'' subscript. The \texttt{simple} algorithm applies an isochrone matched-filter approach in color--magnitude space to identify arcminute-scale spatial overdensities of resolved stars. We used a $\tau =13$~Gyr, $Z=0.0001$ PARSEC-COLIBRI isochrone in the LSST bandpasses \citep{2017ApJ...835...77M} scanned over distance moduli from $16.0\leq(m-M)_0\leq 23.5$ in steps of 0.5 mag. This search was repeated for the $g,r$ and $g,i$ filter pairs, and detections were crossmatched to reduce false positives.

\par
\begin{figure*}[ht!]
    \centering
    \includegraphics[width=\textwidth]{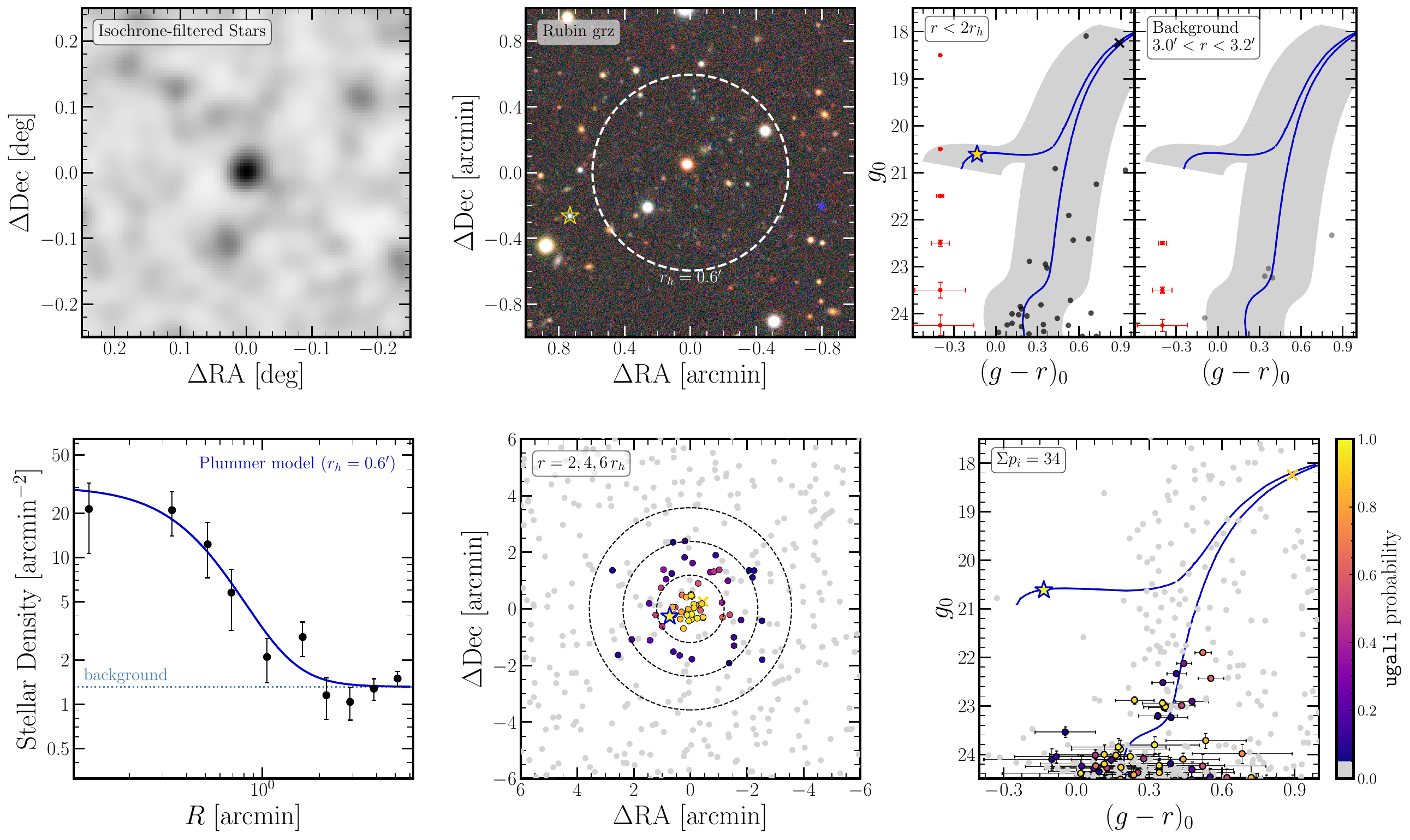}
    \caption{(Top Left) Smoothed spatial distribution of isochrone-filtered stars in a small region surrounding Aquarius~IV. (Top Center) EDP2 color coadd image of a much smaller region centered on Aquarius~IV. (Top Right) CMD of stars located within $2r_h \approx 1.2\arcmin$ of Aquarius~IV (left) and in an equal-area background annulus (right). We highlight the BHB star with a {\large $\star$} due to its importance for our distance estimate. (Bottom Left) Radial density profile of isochrone-filtered stars. The best-fit  Plummer model is shown in blue. (Bottom Center) Spatial distribution of stars in a $12\arcmin \times 12\arcmin $ region centered on Aquarius~IV, colored by \texttt{ugali} membership probability; stars with probabilities $p_i<0.05$ are shown in grey. (Bottom Right) CMD of the same stars shown in the bottom-center panel. The brightest RGB candidate ($g_0 \approx 18$; $\times$ symbol) is ruled out as a member due to its large \textit{Gaia} DR3 proper motion. }
    \label{fig:diagnostic}
\end{figure*}

 Our highest-ranked new detection, Aquarius~IV (Rubin~\Iau), was identified at $8.0\sigma$ in the $g,r$ search and $7.6\sigma$ in the $g,i$ search. The same search also identified all known ultra-faint MW satellites within the EDP2 footprint, including  Sagittarius~II, Aquarius~II, Aquarius~III, and Virgo~III. 
Visual inspection of the coadd images from EDP2 and archival images from the Dark Energy Camera (DECam) revealed an excess of faint, blue, marginally-resolved stars coincident with Aquarius~IV (\figref{diagnostic}), supporting the detection. The identification of a likely blue horizontal branch (BHB) star located $0.8\arcmin$ from the candidate's centroid increased confidence in its reality. 
Retroactive inspection of candidates from a search of DECam data using \texttt{simple} \citep{2026ApJ..1000...87T} revealed a prior detection of Aquarius~IV at  ${\sim}6\sigma$ significance, further supporting its existence.

\section{Morphology and Stellar Population Properties}
We fit the structure and distance of Aquarius~IV with the \texttt{ugali} toolkit \citep{Bechtol15, 2020ApJ...893...47D}. Aquarius~IV's stellar density profile was modeled with a Plummer profile with six free parameters: the centroid coordinates, angular semi-major axis, ellipticity, position angle, and richness (number of stars). Its color–magnitude diagram (CMD) was modeled with the same $\tau=13$~Gyr, $Z=0.0001$ isochrone used in our search, with only the distance modulus left free. For the fit, we refined the stellar sample to $g_0<24.5,r_0<24.5$, $\texttt{refExtendedness}=0$, and $(g - r)_0<1$. Posterior probability distributions for each parameter were derived through MCMC sampling with \texttt{emcee} \citep{2013PASP..125..306F}. We estimated Aquarius~IV's absolute magnitude following \citet{2008ApJ...684.1075M}.
\vspace{.6em}
 \par We report parameter estimates and uncertainties from the posterior medians and 68\% highest density intervals:
\vspace{.4em}
\setlist{nolistsep}
\begin{itemize}[noitemsep]
    \item Centroid coordinates: ($\ra$, $\dec$) = (\Ra\,deg, \Dec\,deg) 
    \item  Azimuthally-averaged half-light radius: \ExtensionRadialArcmin \arcmin,  \PhysicalSizeRadial~\pc
    \item Ellipticity (95\% upper limit): $\epsilon $ \Ellipticity
    \item Distance: $(m-M)_0 =$~\DistanceModulus~(stat.) $\pm$~0.1~(sys.), $D_\odot$ = \Distance
    \item Absolute magnitude: \MvMartin
\end{itemize}
\vspace{.4em}
\par These preliminary estimates require deeper observations to confirm, but they currently place Aquarius~IV on the locus of UFDs in the $M_V$--$r_{1/2}$ plane. Its radius is larger than nearly all MW globular clusters, and we tentatively classify the system as a dwarf galaxy.

\section{Conclusion}
We have presented the discovery of a low-luminosity MW companion, Aquarius~IV, using Rubin EDP2. Aquarius~IV is among the faintest known MW satellites in the distant halo ($D_\odot > 100$~kpc), demonstrating Rubin's ability to reveal ultra-faint satellites at the limits of existing surveys, even in its initial phases of operation. The sensitivity of Rubin LSST will steadily increase, and systems similar to Aquarius~IV are expected to be detectable with $>85\%$ efficiency  using the same search algorithm applied here \citep{2025OJAp....8E..89T}. Rubin LSST thus stands poised to revolutionize the census of ultra-faint MW satellites.

\begin{acknowledgments}
\par WC gratefully acknowledges support from a Gruber Science Fellowship at Yale University. This material is
based on work supported by the National Science
Foundation Graduate Research Fellowship Program under Grant No.\ DGE2139841. This work was supported in part by the National Science Foundation AAG program under Grant No.\ AST-2307126 and AST-2407526.
\par This material is based upon work supported in part by the National Science Foundation through Cooperative Agreements AST-1258333 and AST-2241526 and Cooperative Support Agreements AST-1202910 and 2211468 managed by the Association of Universities for Research in Astronomy (AURA), and the Department of Energy under Contract No. DE-AC02-76SF00515 with the SLAC National Accelerator Laboratory managed by Stanford University. Additional Rubin Observatory funding comes from private donations, grants to universities, and in-kind support from LSST-DA Institutional Members.
\par This publication is based in part on proprietary Rubin Observatory Legacy Survey of Space and Time (LSST) data, and was prepared in accordance with the Rubin Observatory data rights and access policies. All authors of this publication meet the requirements for co-authorship of proprietary LSST data.
\par This research uses services or data provided by the Rubin Science Platform at NSF-DOE Vera C.~Rubin Observatory, which is jointly funded by the U.S.\ National Science Foundation and the U.S.\ Department of Energy, Office of Science.
\end{acknowledgments}

\facilities{Rubin:Simonyi, Rubin:USDAC, Blanco, Gaia}

\bibliography{main}{}
\bibliographystyle{aasjournalv7}

\end{document}